\documentclass[12pt,a4paper]{article}
\usepackage[latin1]{inputenc}
\usepackage{amsmath,graphicx}
\usepackage{amsfonts}
\usepackage{amssymb}
\newcommand{\be}{\begin{equation}}
\newcommand{\ee}{\end{equation}}
\newcommand{\ba}{\begin{eqnarray}}
\newcommand{\ea}{\end{eqnarray}}
\newcommand{\nl}{\nonumber \\}

\begin{document}



\begin{center}
{\Large\bf Illustrative Model for Parity Doubling of Energy Levels}\\
\bigskip
{\bf S. S. Afonin}\\
\smallskip
V. A. Fock Department of Theoretical Physics, St.
Petersburg State University, 1 ul. Ulyanovskaya,
St. Petersburg, 198504,
Russia
\end{center}



\begin{abstract}
A one-dimensional quantum mechanical model possessing mass gap, a
gapless excitation, and an approximate parity doubling of energy
levels is constructed basing on heuristic QCD-inspired arguments.
The model may serve for illustrative purposes in considering the
related dynamical phenomena in particle and nuclear physics.
\end{abstract}


\section{Introduction}

The parity doubling --- the occurrence of almost degenerate
opposite-parity energy levels (or resonances) --- is a widespread
phenomenon in nuclear and particle physics. Thus, a vital
theoretical problem emerges concerning construction of dynamical
models possessing the parity doubling, at least in a certain
regime. The realistic models for this phenomenon are usually quite
involved and output is given commonly by many numbers and plots
which one proposes to compare with the experimental data, see,
e.g., references in a review~\cite{a1}. In such a situation it is
always interesting to develop (over)simplified qualitative models
which, nevertheless, reflect some essential physics such as mass
gap and gapless excitations in the case of the strong
interactions.

In this paper we present an attempt to construct an oversimplified
toy model of this sort. We will propose a one-dimensional solvable
model which can possess an approximate parity degeneracy of energy
levels, with the lowest state --- a gapless excitation --- being a
parity singlet. As long as the orbital motion is impossible in one
space dimension, the definition of parity is drastically different
in this case, it is related to the reflection property of wave
functions. For this reason any one-dimensional model may be
suggestive only, at best. Nevertheless, such models can be
developed towards more realistic schemes or represent some
limiting cases of more realistic models, this makes them rather
curious objects for theoretical analysis.

\section{Formulation of problem}

We will put emphasis on the hadron physics, where the parity
degeneracy of excited states is a typical phenomenon. It is
therefore natural for our discussions to depart from the dynamics
of strong interactions.

It is well known that the lattice data for two static heavy quarks are well fitted by the
Cornell potential,
\be
\label{1}
V(r)\approx -\frac{C\alpha_s}{r}+\sigma r,\quad C>0,
\ee
at all distances available. For light quark systems and for heavy-light quarkonia the
effective confinement potentials are not known, however, an assumption of many models is
that they look somewhat similar.

First of all, for the sake of simplicity we want to "project" the potential in Eq.~\eqref{1} on one space
dimension. It is reasonable to assume that the potential $V(r)$ is then mapped onto some symmetric
one-dimensional potential $V(x)$. Energy levels of any symmetric one-dimensional potential always have a definite
parity equal to $P=(-1)^n$, where $n$ enumerates the energy levels. The even and odd levels alternate,
with the lowest level being always even.

The quantity $\sigma$ in Eq.~\eqref{1} is known to be universal
for the hadron world, it is usually related to the tension of
effective hadron string. On the other hand, this quantity is very
large in comparison with the masses square $m^2$ of unflavored
current quarks in the QCD Lagrangian,
$\sigma\approx(430\,\text{MeV})^2$ vs.
$m^2\approx(7\,\text{MeV})^2$. If we regard $m$ as a "natural"
scale, the slope in $V(x)$ should look then as enormous. On the
other hand, at very short distances the quarks are almost free due
to the asymptotic freedom of QCD. We will mimic the situation by
means of the following picture: at $0<|x|<l$ the slope is zero,
while at $|x|=l$ the slope is infinite. At $r=0$ the Cornell
potential $V(r)$ goes to minus infinity, in one dimension we will
simulate this property by adding the term $-\alpha\delta(x)$,
where the constant $\alpha$ is positive and may be very large, it
will turn out to be the mass gap of the model.

Our suggestions result in a grenade-like potential $V(x)$ displayed in Fig.~1.

In the next section we will show that (a) the lower part of spectrum of potential $V(x)$ reveals
an approximate parity degeneracy of energy levels in the limit
\be
\label{2}
\frac{m\alpha l}{\hbar^2}\gg1,
\ee
and (b) the lowest state is parity singlet.


\begin{figure}
\vspace{-1.5 cm}
\hspace{3 cm}
\includegraphics[scale=0.7]{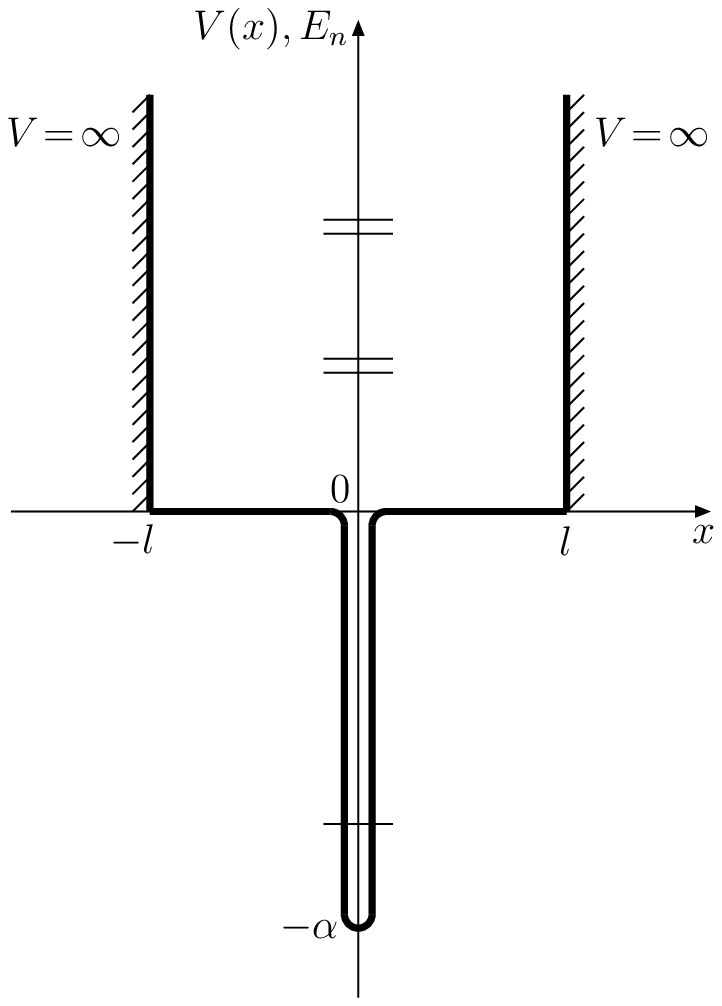}
\vspace{-12 cm}
\caption{The potential $V(x)$ in the constructed illustrative model.
The first energy levels are shown qualitatively.\protect\label{fig1}}
\end{figure}

\section{Solution of problem}

The solutions of one-dimensional Schr\"{o}dinger equation for the potential well and delta-potential
are standard textbook problems in Quantum Mechanics, we will combine them and consider in the limiting
case~\eqref{2}.

The Schr\"{o}dinger equation is
\be
\label{3}
-\frac{\hbar^2}{2m}\Psi''(x)+V(x)\Psi(x)=E\Psi(x).
\ee
As follows from Eq.~\eqref{3}, $\Psi(x)$ is a continuous function at $x=0$, but $\Psi'(x)$ has
to have a discontinuity at
$x=0$ for compensation of the delta-term in the potential $V(x)$
staying in the l.h.s. of Eq.~\eqref{3}.
In order to
obtain the ensuing boundary condition, we integrate Eq.~\eqref{3} in a small interval
around $x=0$,
\ba
0&=&\lim_{\varepsilon\to0}\int\limits_{-\varepsilon}^{\varepsilon}\left(
-\frac{\hbar^2}{2m}\Psi''(x)-\alpha\delta(x)\Psi(x)\right)dx\nl
&=&\lim_{\varepsilon\to0}\left(\Psi'(\varepsilon)-\Psi'(-\varepsilon)+\frac{2m\alpha}{\hbar^2}\Psi(0)\right)
\left(-\frac{\hbar^2}{2m}\right).
\label{4}
\ea
Thus, the boundary conditions at $x=0$ are
\be
\label{5}
\left\{
\begin{aligned}
\Psi'(+0)-\Psi'(-0)&=-\frac{2m\alpha}{\hbar^2}\Psi(0),\\
\Psi(+0)&=\Psi(-0),
\end{aligned}
\right.
\ee
which must be supplemented with
\be
\label{6}
\Psi(\pm l)=0.
\ee

The odd wave functions (w.f.) are not affected by the presence of delta-potential in the central point $x=0$,
the corresponding solution is
\be
\label{7}
\Psi^-(x)=C^-\sin(kx),
\ee
where $C^-$ is a normalization constant, the factor $k$ is
\be
\label{8}
k=\sqrt{\frac{2mE}{\hbar^2}},
\ee
and the spectrum is determined from boundary condition~\eqref{6},
\be
\label{9}
E_n^-=\frac{\hbar^2\pi^2n^2}{2ml^2}, \quad n=1,\,2,\,\dots
\ee

The w.f. of even levels,
\be
\label{10}
\Psi^+(x)=C^+\sin\left(k(|x|-l)\right),
\ee
satisfy boundary condition~\eqref{6} automatically ($k$ is defined in Eq.~\eqref{8}).
The spectrum is now determined from boundary conditions~\eqref{5}, which result in
the following relation,
\be
\label{11}
\tan(kl)=\frac{kl}{\omega},\quad \omega=\frac{m\alpha l}{\hbar^2}.
\ee

One can always choose $\omega\gg1$ such that $kl\ll\omega$ for low enough energy levels.
The quantity in the r.h.s. of Eq.~\eqref{11} is then small, thus, allowing to write
\be
\label{12}
k_nl=\pi n +\epsilon,\quad 0<\epsilon\ll1,\quad \quad n=1,\,2,\,\dots
\ee
From Eqs.~\eqref{11} and~\eqref{12} it follows,
\be
\label{13}
\epsilon\approx\frac{\pi n}{\omega},
\ee
and, hence, the approximate spectrum of even levels is then
\be
\label{14}
E_n^+=\frac{\hbar^2\pi^2n^2}{2ml^2}\left(1+\frac{2}{\omega}\right), \quad n=1,\,2,\,\dots
\ee

It is now easy to see that the odd and even spectra, Eqs.~\eqref{9} and~\eqref{14}, are
approximately degenerate at $\omega\gg1$, which is nothing but the promised statement (a)
(see the definition of $\omega$ in Eq.~\eqref{11} and condition~\eqref{2}).

The obtained solutions exhaust the solutions among trigonometric functions. There exist also
one parity even exponential solution satisfying the imposed boundary conditions. It corresponds to the lowest
energy level, having the following w.f.
\be
\label{15}
\Psi_0^+(x)\approx\frac{\sqrt{m\alpha}}{\hbar}\exp\left(-\frac{m\alpha|x|}{\hbar^2}\right),
\ee
and energy
\be
\label{16} E_0^+\approx-\frac{m\alpha^2}{2\hbar^2}. \ee This
parity unpaired bound state resides in the $\delta$-well at $x=0$.
In the case of the potential $V(x)=-\alpha\delta(x)$,
formulas~\eqref{15} and~\eqref{16} are exact, but the infinite
walls at $x=\pm l$ in Fig.~1 shift slightly upward the lowest
energy level, hence the approximate signs. The existence of this
solution proves statement (b). We note that at $\alpha<0$ this
bound state disappears.

\section{Discussions}

Let us further demonstrate how the proposed toy model may caricature the dynamics
of strong interactions.

First of all, we remark that the property of parity doubling exists only for relatively
low-lying excited levels. It can be shown that for very large $n$, when the opposite
limit takes place in Eq.~\eqref{11}, $kl\gg\omega$, the splitting between the levels
of opposite parity diverges as
\be
\label{17}
E_n^--E_n^+\approx\frac{\hbar^2\pi^2(n-1/4)}{2ml^2}+\frac{\alpha}{2}.
\ee
This property resembles parity doubling in non-strange baryons, where the mass splitting
of parity partners seems to become growing in the highest available excitations~\cite{a1}.

The physical origin of masses of the lightest mesons --- the Goldstone bosons ---
is known to be very different from that of ordinary meson resonances. A similar picture
takes place in the present toy model. First, the ordinary energy levels appear above
$E=0$, i.e. they have the energy gap $\alpha$, while the lowest level is gapless. Second,
the Green function of particle in infinite
potential well is an analytic function of energy in the complex energy plane and
the positions of energy levels are known to coincide with the poles of this Green function,
while the bound state in the $\delta$-well is not such a pole,
which can be seen from the w.f. of this state, Eq.~\eqref{15}, in the momentum representation,
\be
\label{18}
\Phi_0^+(p)\approx\frac{\tilde{C}}{p^2+2m|E_0^+|},\quad \tilde{C}=\frac{m\alpha}{\hbar}
\sqrt{\frac{2m\alpha}{\pi\hbar}}.
\ee

The fact that the energy of the lowest level is proportional to "bare" mass $m$ is suggestive
in relating this state to "would be" Goldstone boson. However, in the limit $m=0$, the finite
discrete spectrum disappears in the model, unless some specific scaling law is imposed. Let us
fix a position of the lowest energy level,
\be
\label{19}
E_0^+\approx-\alpha\xi,\quad 0<\xi\lesssim1.
\ee
Comparing Eqs.~\eqref{16} and~\eqref{19}, we obtain the following scaling law,
\be
\label{20}
m\alpha\approx 2\hbar^2\xi,
\ee
which fixes the w.f.,
\be
\label{21}
\Psi_0^+(x)\approx\sqrt{2\xi}\exp\left(-2|x|\xi\right).
\ee
In particular, at $\xi\approx1$ the lowest level is situated near the bottom of the potential
$V(x)$ ("massless" excitation) and the w.f. gets maximally localized near $x=0$.
The energies of higher states become proportional to the energy gap $\alpha$, {\it e.g.},
the odd energy levels are
\be
\label{22}
E_n^-=\frac{\alpha\pi^2n^2}{4l^2\xi}.
\ee
This scaling has an analogy with the spectra of light mesons, $m_n^2\sim2m_{\rho}^2n$,
where $m_{\rho}^2$ may be also interpreted as a mass gap~\cite{jhep}.

The dynamics of strong interactions leads to a maximal mixing of axial-vector field $A_{\mu}$
and divergence of pseudoscalar field $\partial_{\mu}\pi$, with the corresponding states
$a_1$ and $\pi$ having opposite parities.
The effect results in the symmetric mass splitting,
\be
\label{23}
m^2_{a_1,\pi}\approx m_{\rho}^2\pm m_{\rho}^2,
\ee
in some phenomenological models of chiral symmetry breaking~\cite{jhep,gh,wein}.
A similar phenomenon may take place in the toy model under consideration. The maximal mixing
between $A_{\mu}$ and $\partial_{\mu}\pi$ can be paralleled with the requirement of maximal
value for the overlap integral
\be
\label{24}
\int\limits_0^{\infty}\left|\Psi_1^-(x)\frac{d\Psi_0^+(x)}{dx}\right|dx,
\ee
where the symmetry of potential is taken into account. This requirement is an equation
for the variable $l$ with the parameter $\xi$. Simple calculations yield
\be
\label{25}
\xi\approx y,\quad y\equiv\frac{\pi}{2l}.
\ee
Looking at Eqs.~\eqref{19} and~\eqref{22}, we observe immediately that relation~\eqref{25}
corresponds to minimal possible splitting $E_1^--E_0^+$,
the latter is symmetric with respect to $E=0$,
\be
\label{26}
E_1^-\approx\alpha y,\quad E_0^+\approx-\alpha y.
\ee
It is interesting to note that both in realistic models and in our toy model the centers of
such splittings can be regarded as the position of mass gaps in the underlying theories.

Finally, we would like to notice a conceptual distinction between
the parity doubling mechanism within the considered model and the
effective restoration of broken symmetry in the two-dimensional
quantum-mechanical model of Ref.~\cite{gloz}. The latter
illustrates the hypothetical effective chiral symmetry restoration
in the upper part of hadron spectrum, which is believed to be a
possible reason for the parity doubling in excited light hadrons.
The underlying ideology is very simple --- if Hamiltonian of a
system possesses some symmetry and we introduce an explicit
symmetry breaking term into the Hamiltonian, the low-energy levels
will be perturbed strongly, while the high-energy levels will be
perturbed slightly as if the symmetry were almost unbroken. In
contrast to this scenario, the spectral degeneracy emerges
dynamically within the presented one-dimensional model. The parity
doubling mechanism in the real QCD is not known, it is not
excluded that it has indeed a dynamical origin~\cite{a1}. In
this regard, it should be noted that the spontaneous chiral
symmetry breaking in QCD is a many-body effect, hence, it cannot
be modeled properly in one-particle quantum mechanics because only
explicit symmetry breaking is possible in the latter. In the case
of dynamical symmetries, such an insurmountable difficulty does
not appear since these symmetries reflect internal structure of
the system (see, e.g., discussions in~\cite{a2}) and usually this
internal structure, in some sense, may be projected on lower
dimensions.

\section{Conclusions}

We have proposed a simple toy model which illustrates parity
doubling of energy levels, with the pattern of doubling being
similar to that of in the hadron physics --- the ground state is
parity singlet. It was demonstrated that the model imitates
several phenomena in real hadron physics. Our aim was to construct
the simplest model of parity doubling which nevertheless reflects
some important and very complicated dynamical phenomena in strong
interactions, such as the existence of mass gap and gapless
excitation. Another aim was that the effective parity doubling had
a dynamical origin. Taking into account the claimed purposes, we
hope that our goal is achieved. The model may serve as a base for
complicated extensions which are able already to describe the
related physics quantitatively. For instance, the model proposed
recently by Compean and Kirchbach~\cite{kir} may be regarded as a
nontrivial three-dimensional extension of our toy model.

\section*{Acknowledgments}
The correspondence with M. Kirchbach is gratefully acknowledged.
The work was supported by RFBR, grant 05-02-17477, grant
LSS-5538.2006.2, and by the Ministry of Education of Russian
Federation, grant RNP.2.1.1.1112.

\end{document}